\documentclass[%
twocolumn,
groupedaddress,
nofootinbib,
 amsmath,amssymb,
]{revtex4-2}

\usepackage{xcolor}
\usepackage{amsmath}
\usepackage{mathtools}
\usepackage{amssymb}
\usepackage[utf8]{inputenc}
\usepackage{graphicx}
\usepackage{dcolumn}
\usepackage{bm}

\begin{document}


\title{Irreversibility of symbolic time series: a cautionary tale}

\author{Llu\'is Arola-Fern\'andez}
\email{lluisarolaf@gmail.com}

\affiliation{Instituto de F\'isica Interdisciplinar y Sistemas Complejos IFISC (CSIC-UIB),\\Campus UIB, 07122 Palma de Mallorca, Spain}%

\author{Lucas Lacasa}
\email{lucas@ifisc.uib-csic.es}

\affiliation{Instituto de F\'isica Interdisciplinar y Sistemas Complejos IFISC (CSIC-UIB),\\Campus UIB, 07122 Palma de Mallorca, Spain}%

\date{\today}

\begin{abstract}
Many empirical time series are genuinely symbolic: examples range from link activation patterns in network science, DNA coding or firing patterns in neuroscience to cryptography or combinatorics on words. In some other contexts, the underlying time series is actually real-valued, and symbolization is applied subsequently, as in symbolic dynamics of chaotic systems. Among several time series quantifiers, time series irreversibility --the difference between forward and backward statistics in stationary time series-- is of great relevance. However, the irreversible character of symbolized time series is not always equivalent to the one of the underlying real-valued signal, leading to some misconceptions and confusion on interpretability. Such confusion is even bigger for binary time series --a classical way to encode chaotic trajectories via symbolic dynamics--.
In this article we aim to clarify some usual misconceptions and provide theoretical grounding for the practical analysis --and interpretation--  of time irreversibility in symbolic time series.  We outline sources of irreversibility in stationary symbolic sequences coming from frequency asymmetries of non-palindromic pairs which we enumerate, and prove that binary time series cannot show any irreversibility based on words of length $m<4$, thus discussing the implications and sources of confusion. We also study irreversibility in the context of symbolic dynamics, and clarify why these can be reversible even when the underlying dynamical system is not, such as the case of the fully chaotic logistic map.
\end{abstract}

\maketitle
\section{Introduction}
\begin{figure}[htbp]
\includegraphics[width=1\columnwidth]{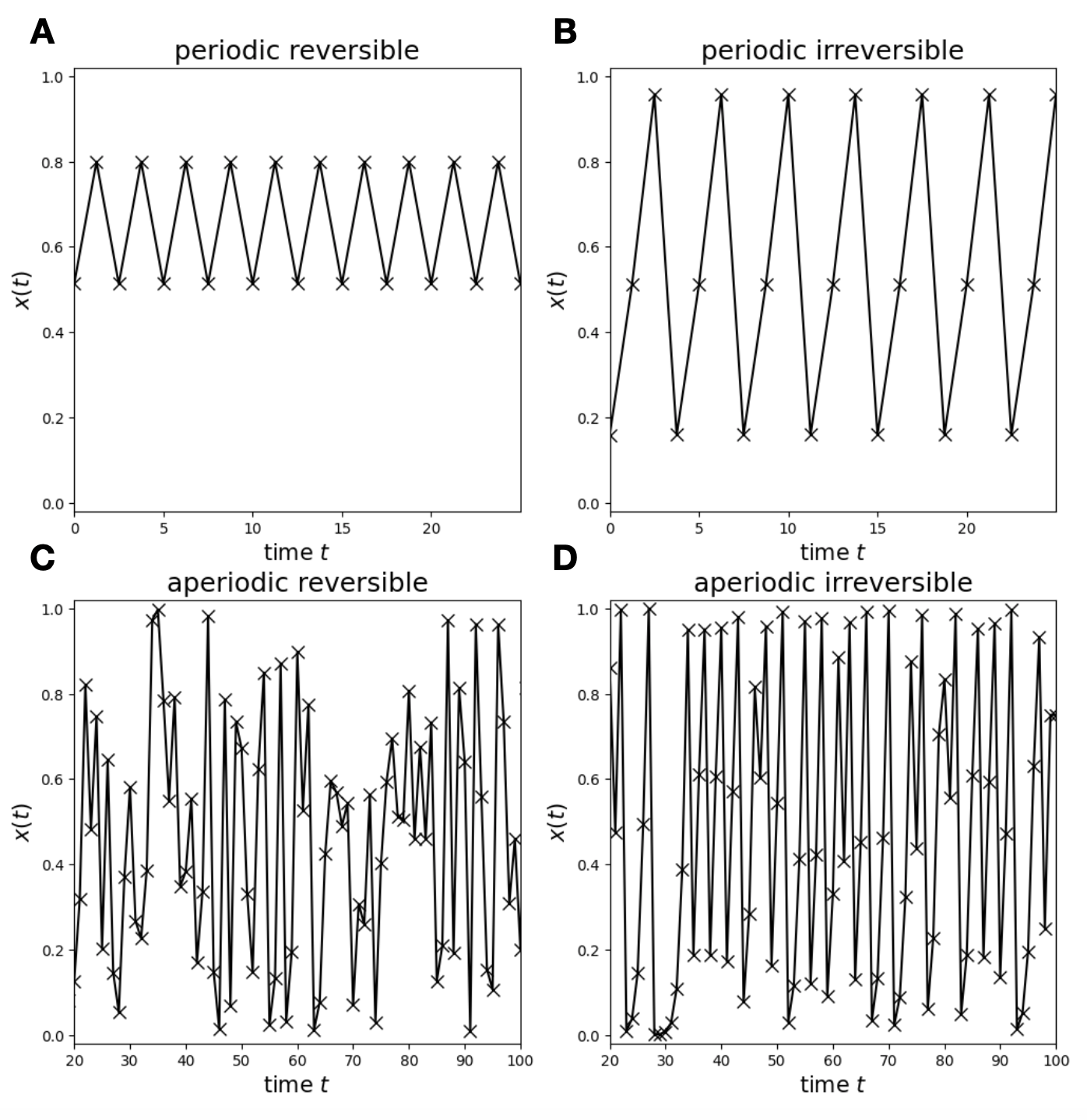}
\caption{Short time windows of reversible (left) and irreversible (right) signals. (Panel A) Periodic reversible trajectory generated from the logistic map $x_{t+1} = rx_t(1-x_t)$ at $r = 3.2$. The signal symmetrically oscillates between two values. (Panel B) Periodic irreversible trajectory of the logistic map at $r = 3.8284$, which corresponds to an island of stability of the map (surrounded by chaos), where the signal oscillates between three values in an asymmetric manner and allows to trivially detect time directionality. (Panel C) Aperiodic reversible trajectory generated from uniform i.i.d. noise. This is an example of a noisy process that is time reversible in the long series limit. (Panel D) Aperiodic irreversible trajectory, generated from the logistic map at $r = 4$. This process is chaotic and dissipative, and therefore irreversible. Trajectories are displayed in a short time window for visualisation purposes, but much longer time series are usually required to correctly estimate time irreversibility in noisy signals.}
\label{Fig_irrev_log0}
\end{figure}

Irreversibility as a close synonym for non-equilibrium thermodynamics is a concept that dates back to the seminal works of Prigogine \cite{prigogine1978time}. Fast-forwarding, statistical time irreversibility, or time irreversibility (TI) in short, is a property of time series (and the underlying processes generating it) whereby the statistics of a time series and its time reversal show significantly different statistical properties (see \cite{zanin2021algorithmic} and references therein). Whereas TI is trivial to observe in non-stationary signals\footnote{See however \cite{lacasa2015time} for a discussion on how to extract meaningful metrics of  TI in non-stationary signals.} or in asymmetric periodic ones, to decide whether a (noisy) stationary signal is reversible (in the statistical sense) is usually nontrivial, see Fig. (\ref{Fig_irrev_log0}) for an illustration.

TI is indeed an important metric and a focus of ongoing research in a variety of fields. In non-equilibrium thermodynamics, the area of stochastic thermodynamics explores the irreversible character of single trajectories \cite{roldan2010estimating}, the onset of apparent violations of the second law, and their quantification via fluctuation theorems \cite{seifert2012stochastic}. In this context, TI is a magnitude that points to the presence of an (entropic) arrow of time, and usually emerges in processes away from equilibrium --where TI can be related to measures of entropy production-- \cite{roldan12}.

Following a similar philosophy, TI has been explored in the realm of (deterministic) dynamical systems, in connection to concepts like phase space contraction and dissipation in chaotic systems \cite{gaspard2005chaos}. Independently, seminal theoretical works on TI appeared in applied statistics, econometrics and time series analysis communities, with classical results on the reversibility of linear stochastic processes or the motto {\it reversibility implies stationarity} \cite{weiss_1975}. Even more independently, in the field of combinatorics on words \cite{lothaire2005applied} the concept of time reversibility is intimately relates to the enumeration of palindrome `words' (blocks of symbols) in abstract sequences of symbols \cite{allouche2003palindrome}.

The fact that advances on apparently the same concept but slightly different definitions have been carried out in different non-communicative disciplines has lead to some confusion. For instance, is a Brownian motion reversible? The answer to this question is not straightforward. One could argue that a Brownian particle in thermal equilibrium is not dissipating and is thus reversible, however the equation that governs Brownian motion (the diffusion equation) is not invariant under time reversal. At the same time, while a Brownian particle trajectory is not on average `producing entropy' and should be catalogued as a reversible process, Brownian motion itself is a non-stationary process, and thus irreversible according to \cite{weiss_1975} (see, however, \cite{lacasa2015time}). Part of the issue might come from conflating the reversible character of a single trajectory with the full (ensemble) process. The latter notwithstanding, while a unification of the polihedric concept of TI awaits, TI --in its different incarnations-- has been widely applied in areas as diverse as financial economics, neuroscience or biological physics,  and we refer the reader to an interesting review article \cite{zanin2021algorithmic} and references therein.

In practice, the dynamical equations (i.e. the Hamiltonian dynamics, the chaotic map or stochastic differential equation) generating the time series is seldom available, and one usually explores TI directly from time series measurements. While measuring TI in real-valued time series (on continuous or discrete time) has been operationalised from a variety of angles \cite{zanin2021algorithmic}, in this paper our aim is to focus on the case where these measurements come in the form of fully discrete time series, i.e. sequences of data where not only time but also the variable itself takes only discrete values. If there is no natural geometric embedding of these values --e.g., if these values are categorical-- then we are dealing with symbolic time series, when the pool of values that can appear in the time series is called an {\it alphabet}. Symbolic time series are indeed ubiquitous across the disciplines, and measuring TI directly from symbolic time series is routinely used in areas such as economics \cite{green1999time, brida2000symbolic}, physiology \cite{cammarota2007time, yao2019quantifying, yao2020permutation}, or genetics \cite{salgado2021time} to name a few.

The main motivation for writing this paper comes from our own experience as researchers and practitioners in TI, where we have encountered several misconceptions and confusions when dealing with TI of such symbolic sequences. For instance, suppose that there exist a (latent) and time irreversible process that generates TI time series, and such series are subsequently symbolized into a sequence of discrete values or symbols. Is the latter symbolic sequence always TI? Unfortunately, this is not always the case, so one needs to be careful (see also the interesting paper \cite{nicolis2011transformation} for a discussion). Our goal is to provide a gentle guidance to some misconceptions and provide theoretical grounding and some new results to guide the researcher and practitioner in the field. 

The rest of the paper is organised as follows: in section II we introduce notation, provide basic definitions and briefly discuss some standard approaches to quantify TI. In section III we establish the two potential sources of irreversibility (that we label Scenario 1 and Scenario 2) that emerge in discrete, symbolic sequences, and we state, prove and discuss a no-go theorem for 3-point statistics that prevents us to conclude the presence (or lack) of TI in binary time series by only looking at 1, 2 or 3-point statistics. Beyond 3-point statistics, in section IV we discuss in detail when Scenario 1 and 2 can and cannot emerge in binary time series generated from deterministic time series via symbolization. This classification shows, for example, that archetypal irreversible processes --such as the fully chaotic logistic map-- are time reversible when symbolized using (standard) generating partitions \cite{beck1995thermodynamics}, whereas they reveal the true TI character of their underlying dynamics when symbolized non-conventionally. At this point we also relate Scenarios 1 and 2 to classic metrics such as the topological entropy or the Kolmogorov-Sinai entropy \cite{beck1995thermodynamics}. Our analysis connects these with the onset of forbidden words in symbolic dynamics and their relation to the area of combinatorics on words \cite{lothaire2005applied}. In Section V we conclude with some take-home messages.

\section{Preamble: notation, definitions and survey of methods}
Throughout this paper, our object of study is a (long) and ordered symbolic sequence ${\cal S}=(x_1,x_2,x_2,\dots,x_n)$, where $x_i$ are symbols extracted from an alphabet $\cal A$ with $|{\cal A}|$ symbols (for the particular case of binary time series, we have ${\cal A}=\{0,1\}$, $|{\cal A}|=2$). We also define a word $w$ of length $m$ as the concatenation of $m \ge 1$  symbols, for instance 0111 is a word of length $m=4$. 

Since we are interested in time series, we assume without loss of generality that ${\cal S}$ is a time sequence, such that we can say $x_t$ is the $t$-th entry of ${\cal S}$ or the entry at time $t$ in an interchangeable way. $\cal{S}$ is said to be stationary if the amount of words of length $1$ is reasonably\footnote{One needs to define a sufficiently large lower bound for the size of the time window over which this check is made} constant over time. We define ${\cal S}$ as the sequence `forward in time', i.e. ${\cal S}\equiv {\cal S}_{\text{fwd}}$, and its time reversed ${\cal S}_{\text{bwd}}=(x_{n},x_{n-1},\dots,x_2,x_1)$. In the limit of large $n$, the hierarchy of normalized frequencies 
$$p(x_1,x_2,\dots,x_m)=\frac{\# \ \text{of occurrences of the word } x_1x_2\dots x_m}{n-m+1},$$
where the denominator takes into account the number of overlapping words in the series, formally defines a stochastic process. A stationary symbolic time series $\cal{S}$ is generally said to be time irreversible --or simply irreversible-- if the statistics of ${\cal S}_{\text{fwd}}$ and ${\cal S}_{\text{bwd}}$ differ. We can formalise this further as follows\\

\noindent {\bf Definition} ($m$-order irreversibility). ${\cal S}$ {\it is irreversible of order $m$ (or simply $m$-irreversible) if
\begin{equation}
p_{\text{fwd}}(w_m) \neq p_{\text{bwd}}(w_m),
\end{equation}
where $w_m$ is the set of all words of length $m$ to be found in the forward and backward time series, respectively.}\\

\noindent We thus say that $\cal S$ is irreversible if there exists a finite $m<n$ for which $\cal S$ is $m$-irreversible (note that $m$-irreversibility implies $m+1$-irreversibility). The lowest value of $m$ for which $\cal S$ is $m$-irreversible is called the intrinsic irreversibility order of $\cal S$.

To operationalise the above definition, we need a way to quantify $m$-irreversibility. A conceptually straightforward way to assess this it to use a similarity metric $\cal D$ between distributions, such that ${\cal D}(p_{\text{fwd}}(w_m),p_{\text{bwd}}(w_m))$ can be computed. Such statistic needs to be unbiased in such a way that for a $m$-reversible sequence ${\cal S}$, we should have ${\cal D}(p_{\text{fwd}}(w_m),p_{\text{bwd}}(w_m))\to 0$ in the limit of large $n$. To assess the significance of ${\cal D}(p_{\text{fwd}}(w_m),p_{\text{bwd}}(w_m))$ for finite $n$, a standard approach is to extract a $p$-value associated to the statistic by shuffling\footnote{Shuffling a sequence preserves its marginal distribution and provides an uncorrelated random sequence which is (asymptotically) statistically reversible} $\cal S$ a number of times, computing the same statistic in each  realization, and checking the percentage of realizations where the measured statistic is larger or equal to the one found for $\cal S$. If such $p$-value is below 0.05, one is confident at the 95\% level that the ${\cal S}$ is $m$-irreversible, and the amount of $m$-irreversibility is given by ${\cal D}(p_{\text{fwd}}(w_m),p_{\text{bwd}}(w_m))$.

Traditionally, authors have used simple measures for $\cal D$ such as the $L_2$ distance \cite{daw00}. Interestingly, if one uses instead the Kullback-Leibler divergence \cite{cover1999elements}, then it turns out that ${\cal D}(p_{\text{fwd}}(w_m),p_{\text{bwd}}(w_m))$ converges, in the limit of large $m$, to a measure of entropy production \cite{roldan12,roldan2010estimating, kawai2007dissipation, parrondo2009entropy}. A different approach that does not require computing $m$-irreversibility and directly aims to estimate differences in the full statistics instead exists. This is based on comparing compression properties of ${\cal S}_{\text{fwd}}$ vs ${\cal S}_{\text{bwd}}$ \cite{kennel2004testing, roldan12}. Other approaches for estimating TI when time series are not necessarily symbolic include the use of multiscale asymmetry \cite{costa2005broken}, visibility graphs \cite{lacasa12, donges2013testing}, or ordinal patterns \cite{martinez2018detection} to cite some (see \cite{zanin2021algorithmic} for a detailed survey).

\section{Sources of $m$-irreversibility in symbolic sequences}

We are interested in finding the words that are potential sources of irreversibility in an arbitrary symbolic time series. Let us consider a large symbolic sequence $\cal S$, with alphabet size $|\cal{A}|$, and let $w_m$ be the set of words of length $m$, which has in principle ${|\cal{A}|}^m$ elements. 

For a given $m$, all words can be classified as palindrome or non-palindrome. A palindrome word is just a word that is invariant under time reversal, e.g. 101 or 01210 are palindromes, whereas 001 or 012 are non-palindromes. The total number of non-palindrome words $\text{NPW}(m,|\cal{A}|)$ can be calculated by subtracting the number of palindrome words from the whole set. Since the number of palindromes is just number of distinct words that can appear in half the word length (if $m$ is even) or $|\cal{A}|$ times the number of distinct words in half the word length minus one (if $m$ is odd), then we get 

\begin{equation}
\text{NPW}(m,|\cal{A}|) =     
    \begin{cases}
     {|\cal{A}|}^{m} - $$|\cal{A}|$$^{(m-1)/2+1} & \text{if} \  m \ odd \\
     {|\cal{A}|}^{m} - $$|\cal{A}|$$^{m/2} & \text{if} \  m \ even
    \end{cases}
    \label{eq:nonpalindromes}
\end{equation}

As an example, take ${|\cal{A}|} = 2$ and $m=3$. The palindrome words are 000, 111, 010 and 101, and the non-palindrome words are 001, 100, 011 and 110. Interestingly, note that all non-palindrome words come in {\it pairs}: one word and its time-reversed, e.g. (001, 100) and (011, 110) in the  preceding example. Thus, observing a particular non-palindrome word in ${\cal S}_{\text{fwd}}$ is equivalent to observing the other element of the non-palindrome pair in ${\cal S}_{\text{bwd}}$. Since by definition palindrome words are invariant under time reversal, any source of irreversibility necessarily emerges only in non-palindrome words, and in particular, comparing the frequencies of both words of each non-palindrome pair.

Furthermore, there are essentially two ways in which these non-palindrome words yield a source of irreversibility, i.e. two scenarios:
\begin{itemize}
    \item Scenario 1 (S1): when in a given non-palindrome pair one word of the pair is present in ${\cal S}$ and the other word is absent (a so-called forbidden word).
    \item Scenario 2 (S2): when in a given non-palindrome pair both words are present in ${\cal S}$ but with different frequencies.
\end{itemize}

\noindent S2 actually englobes S1 as a particular example, but for convenience we keep both scenarios separated. Furthermore, the onset of S1 or S2 predates the practical estimation of  $m$-irreversibility, i.e. there are different ways of actually estimating TI but they all predicate on the existence of S1 or S2. For that reason, the rest of the paper pays special attention to these scenarios.


\subsection{No-go theorem on TI: $m$-irreversibility is impossible for $m<4$ in binary time series}

In real-valued time series, one can find sources of irreversibility already from the inspection of $2$-irreversibility \cite{zanin2021algorithmic}. However, below we show that when dealing with binary time series, this is not the case, and we need at least to go up to words of length $m\geq 4$. Before stating and proving the theorem, we need to introduce some definitions and a useful lemma.\\

\noindent {\bf Definition } (trivial sequence). {\it A binary sequence $\cal S$ is trivial if it eventually becomes constant, i.e.  there exist a $q>0$ for which the sequence ${\cal S}=x_1x_2\dots x_q\overline{0}$ or ${\cal S}=x_1x_2\dots x_q\overline{1}$, where $\overline{s}=sss\dots$. Conversely, a sequence that is not trivial is called a nontrivial sequence.}\\
\

\noindent {\bf Definition } (trivial non-palindrome pair). {\it A binary, non-palindrome pair of words of length $m$ is trivial if one of the words of the pair is either concatenation of a single symbol $\{0\}$ and $(m-1)$ symbols $\{1\}$ or a concatenation of symbol $\{1\}$ and $(m-1)$ symbols $\{0\}$. Conversely, a non-palindrome pair that is not trivial is called a nontrivial non-palindrome pair.}\\

For instance, $(01, 10)$, $(001, 100)$ and $(1000, 0001)$ are trivial non-palindromes pairs of length $m = 2$, $m = 3$ and $m =4$ respectively. Accordingly, the non-palindrome pairs that do not fulfil the above condition are nontrivial, as e.g. $(0100, 0010)$ or $(1010, 0101)$ for $m = 4$ or $(01001,10010)$ for $m=5$, to name a few.\\

The following lemma is also useful to prove the no-go theorem as it discards the words that cannot be sources of irreversibility in binary time series (besides palindrome words, which are trivial reversible sources by definition). \\

\noindent {\bf Lemma } (non-palindrome reversible sources). {\it Let $\cal S$ be a large, nontrivial binary time series of length $n$, and let $z_m$ be the set of all trivial non-palindrome words of length $m < n$. Then} 
$$p_{\text{fwd}}(z_m) = p_{\text{bwd}}(z_m), \ \forall \ m$$ 

\noindent {\it Proof -- } First note that one can always write trivial non-palindrome words either as $a_1a_2\dots a_{m-1}b_m$ or $a_1b_2b_3\dots b_{m}$  where $a_i$ is $\{0\}$ and $b_i$ is $\{1\}$ (or vice-versa) for any $i$ (note that the label here is used to indicate the position of the symbol in the word). 

Let us focus on the first one, and assume that $a_1a_2\dots a_{m-1}b_m$ is present in $\cal S$. For it to occur again, we need to switch from symbol $b$ to `symbol' $a_1a_2\dots a_{m-1}$. This can happen straight away, like $a_1\dots a_{m-1}b_ma_{m+1}\dots a_{2m-1}b_{2m}$, so that the reverse $b_m a_{m+1}\dots a_{2m-1} = b_1a_{2}\dots a_{m}$ already appears in the sequence, or after an arbitrary intermission of size $k$, like $a_1\dots a_{m-1}b_m\overbrace{x_{m+1}\dots x_{m+k}}a_{m+k+1}\dots a_{2m+k-1}b_{2m+k}$, so that the reverse appears at the last $k'$ at which $x_{m+k'} = b$. In any case, for a new $a_1a_2\dots a_{m-1}b_m$ to appear, a $b_1a_2a_3\dots a_{m}$ needs to appear necessarily once, so in the limit  of large sequences, the pair does neither contribute to S1 nor to S2. 

A similar phenomenon happens for $a_1b_2b_3\dots b_{m}$: suppose that we find it in $\cal{S}$ and, after an intermission, we find it again. The intermission can  be (i) void, leading to $a_1b_2b_3\dots b_{m}a_{m+1}b_{m+2}b_{m+3}\dots b_{2m}$ i.e. finding the reverse in the middle; (ii) formed by $a_1x_2x_3\dots x_{k}$, i.e. $a_1b_2\dots b_{m}a_{m+1}x_{m+2}\dots x_{m+k}a_{m+k+1}b_{m+k+2}\dots b_{2m+k}$ which again leads to the appearance of $b_1b_2\dots b_{m-1}a_m$ right after the first $a_1b_2b_3\dots b_{m}$ regardless of $x_2\dots x_{k-1}$, or (iii) formed by $b_1x_2x_3\dots x_k$. In this latter case, the reverse $b_1b_2\dots b_{m-1}a_m$ will necessarily appear as soon\footnote{Interestingly, the size distribution of these intermission blocks could actually have a measurable impact when assessing TI not in terms of the word frequencies but in their recurrence time \cite{salgado2021estimating}, in the event that intermission size follows a complicated non-Markovian process, note however that this might compromise the stationary nature of the time series in the first place.} as any of the $x_i=a$. If none of them take such value (i.e. $x_i = b \ \forall \ i$) then $b_1b_2\dots b_{m-1}a_m$ will appear just after $(m-1)$ appearances of $b$, all of them present in the intermission if $k \geq (m-1)$ or, coming from a concatenation of $(m-1-k)$ symbols $b$ from the initial word and $k$ symbols $b$ from the intermission, if $k < (m-1)$. 

Altogether, in an arbitrary trivial non-palindrome pair, both words have the same frequency, and thus $p_{\text{fwd}}(z_m) = p_{\text{bwd}}(z_m)$, hence concluding the proof. $\square$\\

We are now in the position to state and prove the no-go theorem for binary time series, which restricts the detection of $m-$irreversibility to words of length $m \geq 4$. \\

\noindent {\bf Theorem. } {\it Let $\cal S$ be a large, nontrivial binary time series and let $w_m$ be the set of words of length $m$. Then} 
$$p_{\text{fwd}}(w_m) = p_{\text{bwd}}(w_m), \ \text{for} \ m=1,2,3.$$

\noindent {\it Proof -- } The proof is a constructive one.\\
-- $m=1$: Time reversal leave the marginal distribution of $\cal S$ invariant, and thus trivially we have $p_{\text{fwd}}(w_1) = p_{\text{bwd}}(w_1)$.\\
-- $m=2$: In this case, there is just one non-palindrome pair: $(01,10)$, which is trivial. Using the lemma, we get $p_{\text{fwd}}(01) = p_{\text{fwd}}(10)=p_{\text{bwd}}(01)$. Then we have $p_{\text{fwd}}(w_2) = p_{\text{bwd}}(w_2)$.\\
-- $m=3$: There are a total of $\text{NPW}(3)=4$ non-palindrome words, so two non-palindrome pairs: $(001, 100)$ and $(011, 110)$, which are both trivial. Using again the lemma, we have that, for $m=3$, words have the same frequency for each non-palindrome pairs, and thus $p_{\text{fwd}}(w_3) = p_{\text{bwd}}(w_3)$, concluding the proof. $\square$\\

\noindent This theorem indicates that irreversibility (or lack thereof) in binary time series cannot be concluded on the grounds of $2$- and $3$-irreversibility for binary time series, and words of length $m \geq 4$ need to be considered. To understand this effect, note that for $m = 4$, there are $12$ non-palindrome words but still only $4$ trivial non-palindromes: 0001, 1000, 0111 and 1110 (i.e. two pairs). Thus, the nontrivial non-palindromes can be a source of irreversibility, and these appear only for $m \geq 4$. As an example, consider the short series 0010110010, and note the word 0010, non-palindrome and non-trivial, appears twice while its reversed 0100 does not show up. Instead, for the word 001 to appear again, we observe that its reverse 100 has to show up once in the series. 

Therefore, to correctly estimate the potential sources of irreversibility in binary time series ($|\cal{A}|$ $ = 2$), we shall subtract the two pairs of trivial non-palindromes (for $m \geq 2$) from the set of non-palindrome words. It is then easy to calculate the fraction of words that are potential sources of irreversibility in binary time series, $q(m,|\cal{A}|$ $ = 2)$, with $q(1,2) = q(2,2) = 0$ and  

\begin{equation}
q(m,2) =     
    \begin{cases}
     1-2^{(1-m)/2} -2^{(2-m)} & \text{if} \  m \geq 3 \ odd \\
     1-2^{-m/2} - 2^{(2-m)} & \text{if} \  m \geq 2 \ even
    \end{cases}
    \label{eq:fractions1}
\end{equation}
\noindent Since the number of trivial non-palindromes is always 4 for $m \geq 3$, this correction becomes negligible for larger $m$, but it is the reason behind the impossibility of detecting $m$-irreversibility for $m = 3$. 

\subsection{Symbolic time series beyond the binary case} 

\begin{figure}[htbp]
\includegraphics[width=0.8\columnwidth]{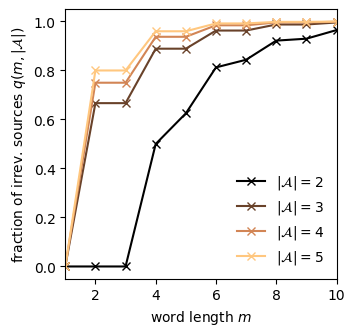}
\caption{Fraction of words that can be sources of irreversibility depending on the word length, for different alphabet sizes. Observe that the black curve, corresponding to a binary alphabet, shows a transition from $q = 0$ to positive values at $m = 4$, as dictated by the no-go theorem, while $q > 0$ for larger alphabets for any $m > 1$.}
\label{Fig_irrev_nogo}
\end{figure}

Observe that for symbolic time series with $|{\cal A}|>2$, all non-palindrome words can contribute to the irreversibility, and the no-go theorem does not apply. As an example, already with $|{\cal A}|=3$ symbols, $2$-irreversibility is possible, e.g. in $01220120122220120$ words of the non-palindromic pair  $(01, 10)$ appear with different frequencies (actually $10$ does not appear, leading to Scenario 1). In this case, we can then use directly the formula given by Eq. (\ref{eq:nonpalindromes}), normalized by $|{\cal A}|$, to estimate the fraction of potential irreversible sources, leading to
\begin{equation}
q(m,|{\cal{A}}| > 2) =     
    \begin{cases}
     1-|{\cal{A}}|^{(1-m)/2} & \text{if} \  m \ odd \\
     1-|{\cal{A}}|^{-m/2} & \text{if} \  m \ even
    \end{cases}
    \label{eq:fractions2}
\end{equation}
In Fig. (\ref{Fig_irrev_nogo}), we plot the fraction of irreversibility sources depending on word length $m$ for several alphabet sizes, computed directly from Eq. (\ref{eq:fractions1}) and Eq. (\ref{eq:fractions2}). We observe that for $|{\cal{A}}| = 2$, the fraction is $q > 0$ for $m \geq 4$, while for $|{\cal{A}}| > 2$, we can detect $m$-irreversibility even for $m = 2$.

In retrospect, the absence of the no-go theorem might be (inadvertently) at the core of TI quantification methods proposed for symbolized time series, such as in \cite{cammarota2007time} where the authors directly propose a ternary coding ($|{\cal A}|=3$), in \cite{salgado2021estimating} where the authors consider a 3-state Markov chain and look into recurrence times,  or in \cite{daw00}, where the authors state (without further discussion) that `we find it
desirable to use larger alphabet sizes [...], in contrast to the usual binary alphabets [...]'. Incidentally, in Panel B, Fig. (3) of \cite{daw00}, the authors seem to find $2$-irreversibility and $3$-irreversibility for $|{\cal A}|=2$. This is in violation of the theorem proved above, and we thus conjecture that such figure might have a typo somewhere (actually, in Panel A Fig. (3) of \cite{daw00} the authors correctly find the lack of $3$-irreversibility for $|{\cal A}|=2$, contradicting Panel B).


\subsection{Further remarks}

Observe that, in practice, the statistics of word frequencies for large length $m$ will be poor when the series size $n$ is not large enough. These finite-size effects make the no-go theorem even more problematic in practice, as the length of the sequence which is required to accurately estimate $m$-irreversibility grows exponentially with $m$, so TI is difficult to estimate in (short) binary time series, which is typically the case in experimental observations. If additional information of the process is eventually available (e.g. if the process runs in continuous time, and there is metadata on the time spent between transitions between states), then alternative quantifications of TI have been proposed \cite{harunari2022learn}. Unfortunately, many empirical time series are plainly symbolic, they are not the outcome of a sort of telegraph process with latent time statistics. Interestingly, recurrence-time statistics might then be used instead \cite{salgado2021estimating}. \\
\noindent For instance, one could assess the number of symbols between two events of $001$, and compare these statistics with the recurrence time of $100$. When imposing complicated non-Markovian processes for  the size of the intermissions,  then one might actually get different recurrence-time statistics, e.g. for a sequence
\begin{gather}
{\cal S}={\bf 001}\underbrace{10101010}_{\tau_1}{\bf 100}\underbrace{000\dots00}_{\tau_2}{\bf 00 1}\underbrace{10101101\dots}_{\tau_3} \nonumber \\ 
{\bf 100}\underbrace{000\dots{ 00}}_{\tau_4}{\bf 001}\underbrace{11101101\dots{ }}_{\tau_5}{\bf 100}\dots,
\end{gather}
then the recurrence time for $001$ is $(\tau_1+\tau_2+3, \tau_3+\tau_4+3, \tau_5+\tau_6+3,\dots)=(\tau_i + \tau_{i+1}+3)_{i=1}^{\dots}$ whereas the one for $100$ would be 
$(\tau_2+\tau_3+3, \tau_4+\tau_5+3, \tau_6+\tau_7+3,\dots)=(\tau_i + \tau_{i+1}+3)_{i=2}^{\dots}$, so statistically the same recurrence times for stationary intermissions --where the index $i$ can be dropped-- and different ones for time-dependent (i.e. non-stationary) intermissions: finding non-trivial patterns might therefore compromise the stationarity of the signal in the first place.

In any case, the no-go theorem clarifies why quantifying irreversibility is harder in binary time series than in real-valued ones. When symbolizing time series, practitioners usually have a bias towards binary time series as these are simpler and require smaller time series size to get accurate word statistics, hence potentially falling into the domain of the no-go theorem inadvertently. Moving from binary to symbolic sequences with a larger alphabet indeed allows us to circumvent the no-go theorem. 

Furthermore, we reckon that researchers also tend to overuse binary time series due in part to the celebrity of information theory or symbolic dynamics theory (see below). For instance, symbolic dynamics elegantly shows that many chaotic properties can be retrieved from suitably binarized time series of chaotic trajectories. Unfortunately, time irreversibility is not consistently one of these properties, because the motion on this symbolized phase space can appear to be strictly random \cite{grassberger1988symbolic} --and thus, time reversible-- as we discuss in the following lines.

\section{Irreversibility in and from symbolic dynamics}
Binary time series are classically found in the mathematical field of symbolic dynamics \cite{beck1995thermodynamics}, where an interval map $x_{t+1}=f(x_t;r), \  x\in[a,b]$ is symbolised by defining a {\it partition} of the interval $[a,b]=\bigcup_{i=1}^k [a_i,b_i)$. The case $k=2$ symbols has received a lot of attention, mainly due to the fact that unimodal maps --canonical low-dimensional examples of chaotic motion-- typically have so-called generating partitions \cite{beck1995thermodynamics} with two symbols: such is the case for e.g. the chaotic logistic map  $x_{t+1}=rx_t(1-x_t), \ x \in [0,1]$, whose binary partition $[0,1]=[0,1/2)\cup[1/2,1]$ induces a symbolic dynamics --binary time series-- from which important properties can be retrieved  \cite{james2014chaos}, for instance, the  time series generated from an initial condition $x$ which is iterated via $f$ and later symbolized uniquely defines $x$.

Chaotic time series are of relevance in the study of irreversibility, as it is often the case that chaotic dissipative maps yield irreversible dynamics, a property that is easily retrieved from the statistics  of the real-valued time series \cite{zanin2021algorithmic}. But is irreversibility conserved in the coarse-grained description offered by the symbolic dynamics? To discuss such question, let us consider two important and classic measures. The so-called topological entropy $K(0)$ counts the number of possible words of length $m$ that are actually present in $\cal S$, such that
\begin{equation}
    K(0)=\lim_{m \to \infty}\frac{1}{m}\ln(|w_{m}|).
\end{equation}
(rigorously, $K(0)$ indeed converges to the topological entropy only when the phase space partition is generating \cite{beck1995thermodynamics, lacasa2018dynamical}). It is straightforward to show that, for $|{\cal A}|=2$ symbols, then onset of S1 directly implies $K(0)<\ln(2)$.

Another important result from the theory of symbolic dynamics is that when $f$ is a chaotic map, its (positive) Lyapunov exponent $\lambda$ can be estimated from the so-called block entropy rate
\begin{equation}
    h_m = \frac{-1}{m}\sum_{w_m} p(w_m) \log p(w_m), 
    \label{Eq:Hm}
\end{equation}
via the so-called Pesin identity \cite{beck1995thermodynamics}, that identifies $\lambda$ with the Kolmogorov-Sinai entropy $h_{KS}$ of the system, where
\begin{equation}
    h_{KS}=\lim_{m\to \infty} h_m
    \label{eq:ks}
\end{equation}
if the partition used to symbolise the trajectory is again a so-called {\it generating partition} (in general, a suitable optimization over all possible partitions needs to be made). Observe that Eq. (\ref{Eq:Hm}) goes above and beyond the topological entropy as it is indeed looking into the frequencies (the measure) of words in ${\cal S}$, not only their existence. In that sense, for $|{\cal A}|=2$ symbols, it is easy to see that onset of Scenario 2 would imply $h_{KS}<\ln(2)$. 
Note at this point that $h_{KS}$ is not a standard entropy as it is not an extensive variable, it is actually an entropy rate. As a matter of fact, Eq. (\ref{eq:ks}) can also be applied for symbolic sequences that are not necessarily symbolizations of deterministic trajectories (e.g. stochastic  processes, and in general any kind of empirical symbolic sequence), and in that event, Eq. (\ref{eq:ks}) converges to the entropy rate of the sequence $\cal S$, or equivalently to the entropy rate of the stochastic process that generated $\cal S$ \cite{cover1999elements} (not to be mistaken with the Shannon entropy of a discrete random variable with support $\cal A$, which essentially is $h_1$).

It is well known that, depending on the specific partition employed and the map's parameter $r$, then the resulting symbolic sequence of a chaotic map can have (i) no forbidden words (whose frequencies could in principle be equal --leading to lack of irreversibility, and a constant $h_m=\ln |\cal A|$-- or different, leading to Scenario 2), or (ii) a hierarchy of forbidden words (a so-called grammar \cite{beck1995thermodynamics, grassberger1988symbolic}), leading to  Scenario 1.\\

\begin{figure*}[!t]
\includegraphics[scale=0.17]{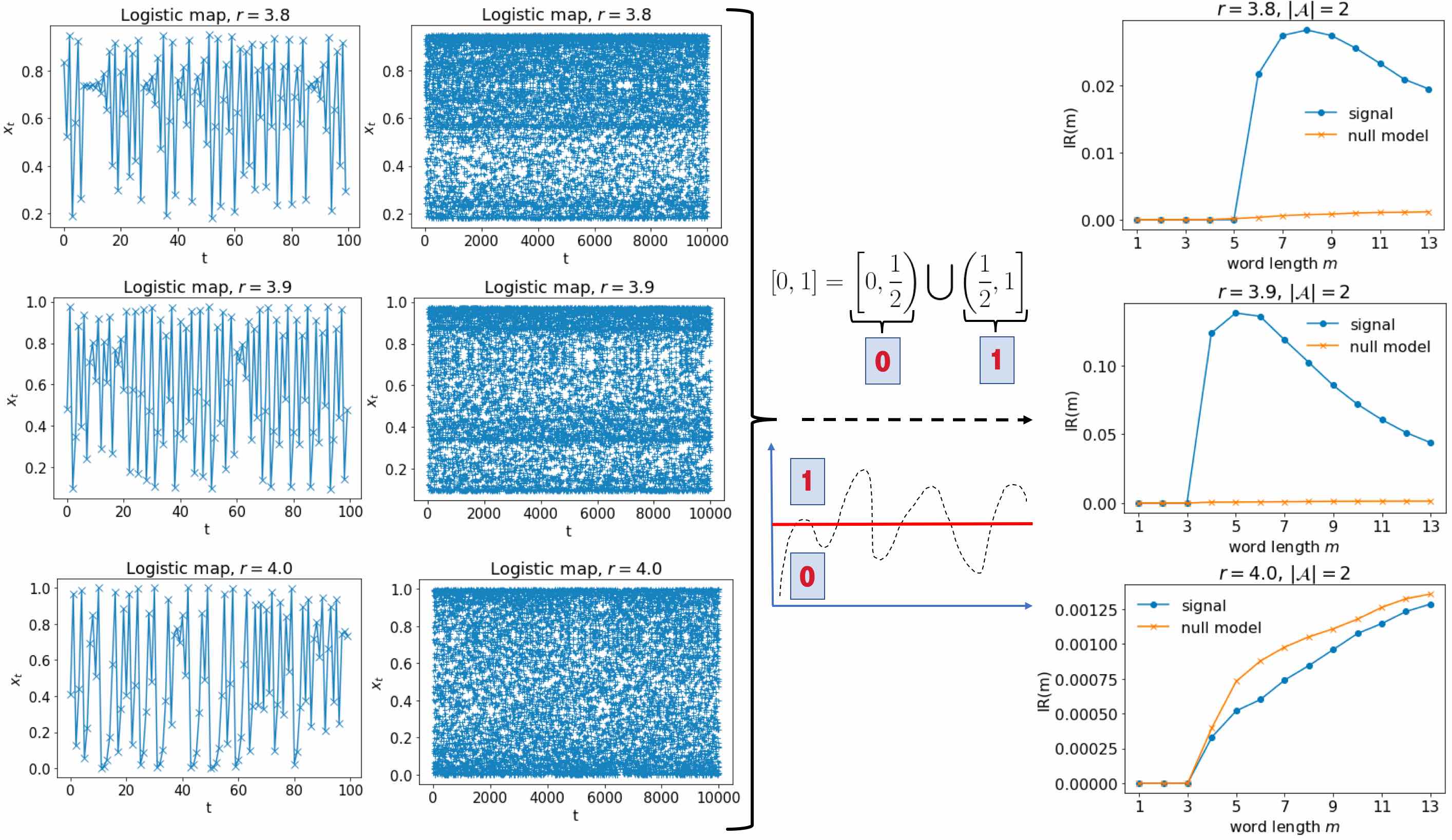}
\caption{Illustration of the resulting $m$-irreversibility (as computed using the $L_2$ norm) for binary time series of $10^5$ data points, generated from the logistic map $x_{t+1}=rx_t(1-x_t)$ for three different values of $r$ where the map is chaotic: $r=3.8$ (Pomeau-Maneville scenario), $r=3.9$ and $r=4.0$ (fully chaotic and topologically conjugate to the Bernoulli shift). In every case the time series are symbolised via a 2-symbol partition  $[0,1]=[0,1/2)\cup (1/2,0]$. Such partition is generating only for $r=4.0$ --and in that case the induced symbolic dynamics is a conventional Markov chain, i.e. time reversible-- and for other values $r\neq 4$ is expected to induce a complicated non-Markovian process, with the onset of different frequencies for some non-palindrome pairs (inducing $m$-irreversibility via Scenario 2), and the onset of forbidden words (grammar), leading also to $m$-irreversibility via Scenario 1. The last column shows both the results of $m$-irreversibility as computed from the symbolized time series and the same value applied to a null model where the symbolized time series has been shuffled (such null model by construction is time reversible for all $m$, and thus any finite value is just a finite size effect). In every case, we find lack of irreversibility for  $m<4$, as expected by virtue of the no-go theorem. For $r=4.0$ (bottom panel of the last column), results for $m\geq 4$ are not different from the null model, thus reversibility cannot be rejected, as expected. For $r=3.8$ and $r=3.9$, we find signals of $m$-irreversibility starting from $m=4$ for $r=3.9$ and from $m=6$ for $r=3.8$, highlighting that each case possess different grammars and the induced This is non-Markovian process, while irreversible, shows different degrees of irreversibility.}
\label{Fig_irrev_log}
\end{figure*}

\begin{figure*}[htbp]
\includegraphics[width=1.68\columnwidth]{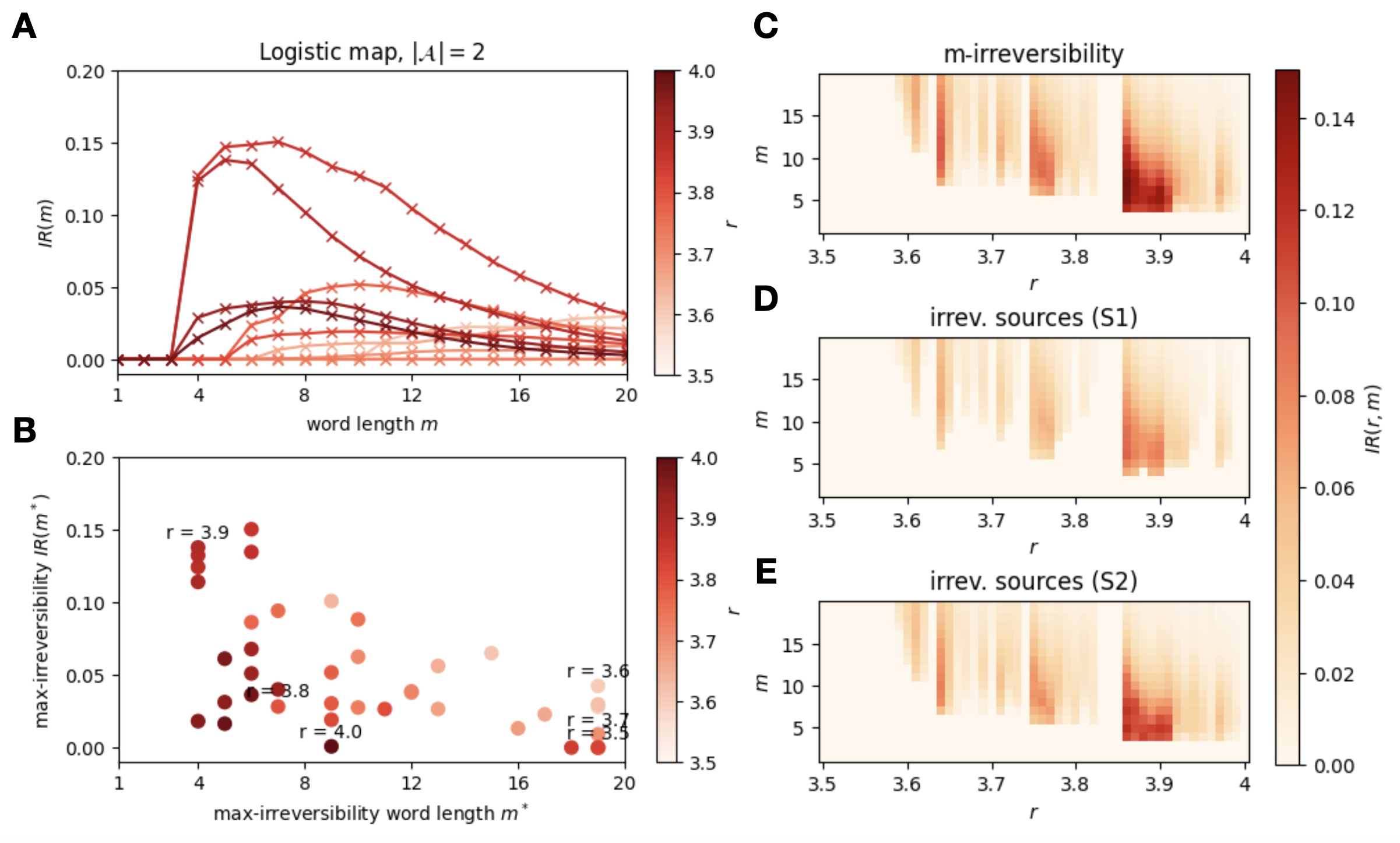}
\caption{(Panel A) Curves of $m$-irreversibility IR($m$) (computed via $L_2$ norm) for binary time series of $3\cdot 10^6$ data points, extracted from the logistic map $x_{t+1}=rx_t(1-x_t)$ at different values of $r$. (Panel B) Scatter plot of the maximum value of $m$-irreversibility IR($m^*$) as a function of the value of the word length $m^*$ where such value is found. We can see that irreversibility curves are systematically non-monotonic, albeit with a peak location $m^*$ which varies non-trivially for different values of the map's parameter $r$, and large irreversibility values are typically related to $m$-irreversibility curves that peak at $m^*\in [5,10]$. (Panel C) Heatmap of the $m$-irreversibility value depending on $r$ and $m$ in the logistic map. (Panel D) Heatmap of the $m$-irreversibility computed only for the words satisfying Scenario 1 (S1). (Panel E) Heatmap of the $m$-irreversibility computed only for the words satisfying Scenario 2 (S2), neglecting in the sum of Eq. (\ref{eq:IRm}) the words that contribute to S1.}
\label{Fig_irrev_log2}
\end{figure*}

\noindent {\bf Sequences that lack forbidden words}. By topological conjugacy, any map which in conjugate to the Bernoulli shift lacks forbidden symbols in its symbolic dynamics \cite{beck1995thermodynamics}. This is the case for the tent map or the Ulam map $x_{t+1}=1-2x_t^2$, equivalent to the fully chaotic logistic map ($x_{t+1}=rx_t(1-x_t)$ for $r=4$). In all these cases, with a binary partition $|{\cal A}|=2$ which is generating, the number of possible (admissible) words of length $m$ is indeed $|w_{m}|=2^{m}$.  The topological entropy is then $K(0)=\ln(2)$: Scenario 1 does not take place. Interestingly, the KS entropy entropy also equals $\ln 2$ (the Lyapunov exponent of the map). This in turn suggests that $p(w_{m})\to 1/2^{m}$, i.e. symbolic (binary) series might be reversible for such chaotic maps (even if the real-value trajectories of these maps are easily seen to be irreversible, see for instance \cite{lacasa12}). As a matter of fact, for $x_{t+1}=4x_t(1-x_t)$  the usual partition is not only generating, it is also a Markov partition (that yields ${\cal S}$ a topological Markov chain). Not only that, such partition makes ${\cal S}$ a conventional Markov chain, which is trivially reversible. This result is in stark contrast with the fact that the real-valued map at $r=4$ actually generate irreversible trajectories. For other values of $r$, the partition does not generate a conventional Markov chain.

\noindent {\bf Sequences with forbidden words}. 
If the partition is not generating, or if the map is not conjugate to Bernoulli, the results above don't hold.
Indeed, in the logistic map outside $r=4$ and using the usual partition $[0,1/2)\cup(1/2,1]$ (which is again generating) leads to a stochastic process associated to the statistics of $\cal S$ which in general is non-Markovian \cite{beck1995thermodynamics}. Interestingly, it turns out that for $r\neq 4$ a whole set of forbidden words emerge. Such set is different for each value of $r$ and is the so-called {\it grammar} \cite{beck1995thermodynamics, grassberger1988symbolic}. The existence of forbidden words of size $m$ inside such grammar that affect single words of each non-palindrome pair (i.e., for instance, having 0101 as a forbidden word but not 1010) is a first guarantee of finding $m$-irreversibility, according to Scenario 1.\\

\noindent To gain some intuition and further support the claims above, we now run some numerical simulations where we track $m$-irreversibility in a number of situations (note that along this paper we provide a Python code to test $m$-irreversibility). For the sake of exposition, here we choose a simple way to compute $m$-irreversibility IR($m$) in terms of the $L_2$ distance between respective distributions \cite{daw00}
\begin{equation}
    \text{IR}(m) = \sqrt{\sum_{w \in w_m} [p_{\text{fwd}}(w) - p_{\text{bwd}}(w)]^2 },
    \label{eq:IRm}
\end{equation}
but one can expect the same qualitative results if another metric to measure the distinguishability between the distributions is used (as e.g. the Kullback-Leibler divergence or the Jensen-Shannon distance \cite{roldan12,zunino22}).

In Fig. (\ref{Fig_irrev_log}) we depict IR($m$) for symbolized time series (with $|{\cal A}|=2$ symbols using a generating partition) from the logistic map $x_{t+1}=rx_t(1-x_t)$, for several values of $r$ where we know that the map is chaotic, dissipative, and time irreversible. We find that for $r=4$ the resulting binary time series is indeed $m$-reversible at all $m$, as expected, even if the non-symbolized one is clearly irreversible \cite{lacasa12}, due to the fact that the precise partition of the phase space used to symbolize the time series is inducing a conventional Markov chain in the symbolized sequence. For other values of $r$, a grammar emerges, and there are clear signs of $m$-irreversibility  (starting at $m>3$ as per the no-go theorem). Note that the grammar of forbidden words and the specific frequencies of words in each non-palindrome pair non-trivially depends on the specific value of $r$, leading to very different net values of $m$-irreversibility (albeit all positive and significant as they deviate from the results of a null model): for instance for $r=3.9$, IR($m$) is non-zero starting from $m=4$, and peaks at $m=5$ with a value around 0.14, whereas for $r=3.8$, IR($m$) is non-zero starting from a larger $m=6$, the peak is located at $m=8$ and the value of irreversibility is substantially smaller.

\begin{figure*}[htbp]
\includegraphics[width=1.68\columnwidth]{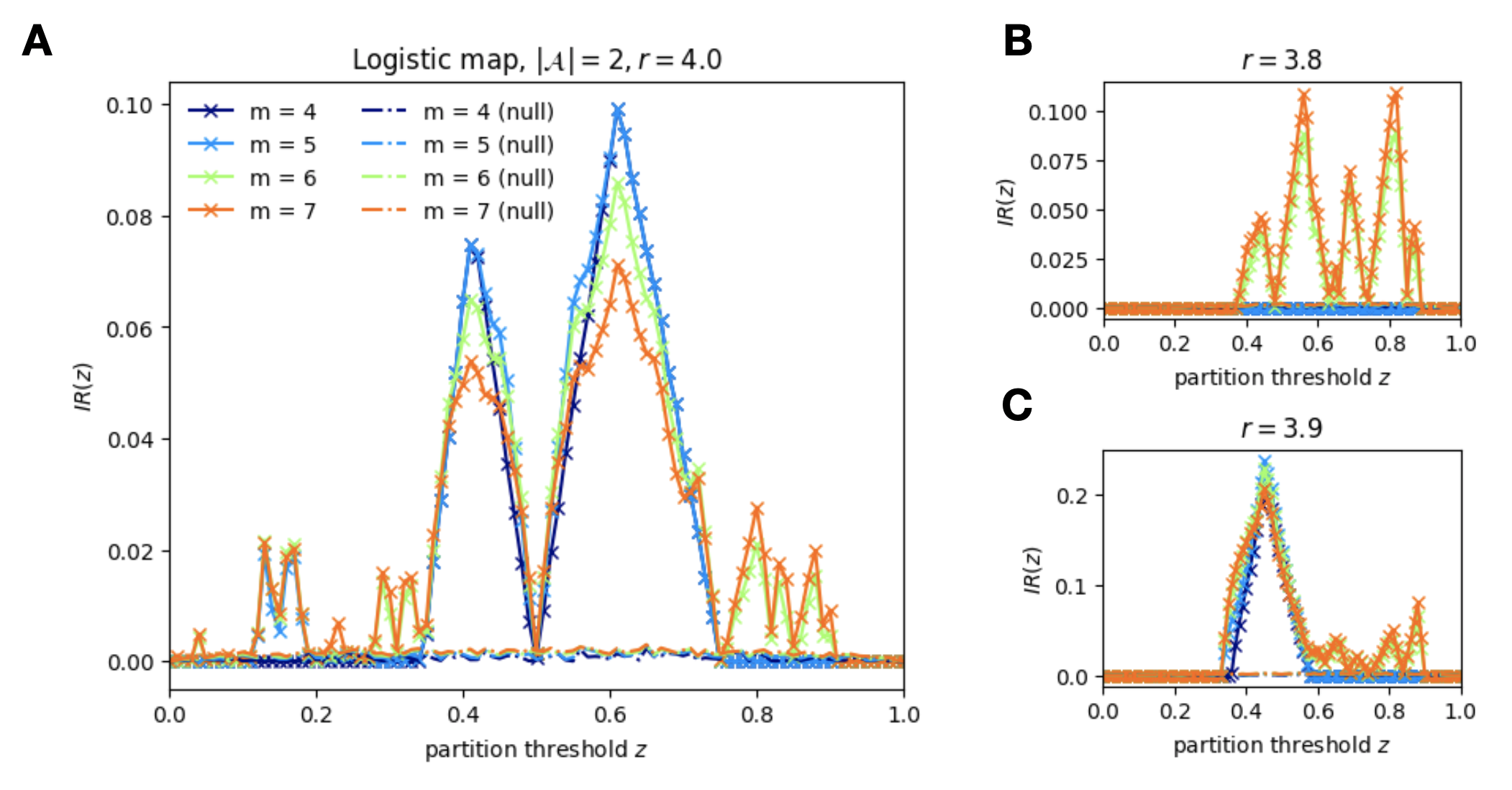}
\caption{(Panel) A. $m$-irreversibility, computed using Eq. (\ref{eq:IRm}), depending on the partition threshold of the binarisation process in the logistic map with $r = 4$ and for several values of the word length $m$, in time series of $3 \cdot 10^6$ points. The dashed lines correspond to the irreversibility of the null-model, where the symbols have been randomly shuffled, and no irreversibility is detected. Panel B (and C) shows the results for $r = 3.8$ (and $r = 3.9$).}
\label{Fig_irrev_log3}
\end{figure*}

These findings are further illustrated in Fig. (\ref{Fig_irrev_log2}), where we scan a wide range of $r$ values from $3.5$ to $4$. This range spans the region where chaotic behavior occurs in the logistic map. In fact, only the chaotic region is our matter of interest here, because we aim at detecting irreversibility in noisy stationary signals. For this reason, in Fig. (\ref{Fig_irrev_log2}) we neglect the contributions to $m$-irreversibility produced by the purely periodic, asymmetric motion that occurs on islands of stability, such as the one arising around $1 + \sqrt{8} \approx 3.83$ \cite{zanin2021algorithmic}. On this island, irreversibility is much larger than the one measured at any other value of $r$, but this large magnitude comes from a pure periodic motion, where the dynamics shows asymmetric oscillations among a few values of $x$. Periodic irreversibility is trivial to detect (see also Fig. (\ref{Fig_irrev_log0}) for more details), and thus here we focus only on irreversibility sources coming from noisy (here chaotic) stationary signals. 

We can now proceed to discuss in more detail Fig. (\ref{Fig_irrev_log2}). In Panels A and B, we observe that, whereas we find that IR($m$) is typically non-monotonic where the location of maximum irreversibility $m^*$ depends non-trivially on $r$, those values of $r$ for which the symbolic time series has consistently larger irreversibility coincides with irreversibility peaking at word lengths $m^* \in [5,10]$. In Panel C, we use a heatmap to visualize how our irreversibility measure depends on both $r$ and $m$ in the logistic map using the binary partition $[0,1]=[0,1/2)\cup[1/2,1]$. One can observe that irreversibility increases from zero to some positive value around the onset of chaos in the map, at $r \approx 3.57$ and its presence is sustained across most of the $r$ values in the range. Besides the aforementioned island of stability around $r \approx 3.83$, we observe other chaotic regions where irreversibility is negligible, as e.g. at the well-studied $r = 4$ due to the Markovian nature of the simbolised dynamics under the symmetric partition. 

In Panel C (D), we plot the contribution of the terms corresponding to the S1 (S2) sources of irreversibility in Eq. (\ref{eq:IRm}) to illustrate the effect of the grammar (the presence of forbidden words) in the binarised logistic map. We observe a strong correlation between the presence of both sources of irreversibility, showing that a dynamics with forbidden words (S1) will usually have strong asymmetry between the frequencies of appearance in the forward and backward series even for the set of non-forbidden words (S2). This effect is well observed, for instance, around $r \approx 3.9$, where $m$-irreversibility is larger at a very low value of word length $(m^* = 4)$. In this region, S1 indeed occurs, but the most part of the $m$-irreversibility comes from S2. 
Overall, the heterogeneity in the heatmaps highlights the nontrivial dependence of the grammar of the map on the chosen values of $r$ and $m$, even when we are using the symmetric partition $[0,1]=[0,1/2)\cup[1/2,1]$ that is known to be a generating partition for any value of $r$.

In Fig. (\ref{Fig_irrev_log3}), we explore the behavior of $m$-irreversibility in the logistic map for non-generating partitions. We vary the partition threshold $z$ such that $[0,1]=[0,z)\cup[z,1]$ (only $z=0.5$ leads to a generating partition). In Panel A, we observe the non-monotonous behavior of $m$-irreversibility depending on $z$, when the parameter is set to $r = 4$, for several values of $m$. A valley with IR(m) $ = 0$ at $z = 0.5$ is surrounded by two peaks of large irreversibility. In particular, the maximum of irreversibility occurs at $z = 0.6$, surprisingly indicating that a binary dynamics generated from a non-generating partition of the unit interval is the one that maximizes the $m$-irreversibility in this map. However, as observed in Panels B and C, if the parameter of the map changes, the whole curve and the location of the maxima also change, indicating that the irreversibility of the process is highly dependent, not only on the value of $r$ and the word length $m$, but also on the partition that is selected.


\section{Discussion}
When dealing with (binary or symbolic) time series, results on TI should be taken with caution. In particular, false negatives can occur as we have explained. We summarise here a number of take-home messages:

\emph{i)} The statistics of words of length $m<4$ are equivalent for binary time series and its time-reversed as shown by the no-go theorem proved in Section III. As a result, Time Irreversibility (TI) cannot be concluded on such basis, as statistics of longer words need to be considered. This limitation is removed when using an alphabet of more than two symbols, so in practice moving away from binary to ternary (or above) time series is preferred, even if this increase in the alphabet size requires longer time series for a correct estimation of word statistics. This no-go theorem explains why current algorithmic proposals to assess TI in symbolic time series heuristically move away from alphabets with only two symbols, despite the fact that researchers have a positive bias towards binary time series, usually related from the celebrity of information theory and symbolic dynamics.

\emph{ii)} Some time irreversible processes such as chaotic dissipative processes can be misleadingly seen as time reversible when symbolized, this is for instance the case of the fully chaotic logistic map $x_{t+1}=4x_t(1-x_t)$ symbolized via the generating partition $[0,1]=[0,1/2)\cup(1/2,1]$ (binary symbolization), which induces a conventional Markov chain in the symbolic dynamics. Non-conventional partitions (e.g. non generating binary partitions) induce complicated non-Markovian stochastic processes in the symbolic dynamics from which time irreversibility is restablished and can then be correctly estimated (still, from words of length $m\geq 4$). In the absence of a clear criterion for an adequate partition, to assess TI directly from the real-valued time series is preferred, and otherwise, a partition with more than two symbols is a cautious choice.

\emph{iii)} The onset (or lackof) time irreversibility from symbolic dynamics (i.e. for (real-valued) deterministic maps whose time series are symbolized using a suitable partition of the phase space) is related to the emergence of forbidden words (the grammar) and/or the emergence of asymmetries in the frequencies of words for each non-palindromic pair. The amount of such non-palindromic pairs grows exponentially with the length of the word. The grammar is very dependent on the partition and the parameters of the underlying dynamical system, and a case-by-case analysis needs to be applied.

\emph{iv)} Besides the statistics of words, irreversibility can also show in the statistics of word recurrence times. For continuous-time discrete state statistics, one simply quantify the two source of irreversibility: statistics of occurrence of different non-palindrome words, and waiting time statistics \cite{harunari2022learn}. For fully discrete time series with no latent information on the time between state transitions, such information is directly encoded in the time of the intermissions, and thus one can retrieve recurrence times between words accordingly \cite{salgado2021estimating}. However, finding source of irreversibility in those recurrence times might require that the symbolic time series to be non-stationary, what poses a conceptual problem in the first place.\\

\noindent {\bf Acknowledgments:} The authors would like to thank E. Rold\'an and M. Zanin for useful discussions. We acknowledge funding from project DYNDEEP (EUR2021-122007) from the Agencia Estatal de Investigaci\' on. LL additionally acknowledges funding from project MISLAND (PID2020-114324GB-C22), and the María de Maeztu project CEX2021-001164-M.\\

\noindent {\bf Code availability:} Upon publication, Python codes of $m$-irreversibility will be available in a Github.

\bibliography{apssamp}

\end{document}